\documentclass[10pt,aps,prd,floatfix,twocolumn,nofootinbib,superscriptaddress,preprintnumbers]{revtex4-2}
\pdfoutput=1

\usepackage{graphicx} 
\usepackage{amsmath,amssymb,bm}
\usepackage{xcolor}
\usepackage[colorlinks,allcolors=blue]{hyperref} 

\begin{document}

\preprint{TU-1241}

\title{Isotropic cosmic birefringence from an oscillating axion-like field}
\author{Kai Murai}
\affiliation{Department of Physics, Tohoku University, Sendai, Miyagi 980-8578, Japan}

\begin{abstract}
We propose a new mechanism for isotropic cosmic birefringence with an axion-like field that rapidly oscillates during the recombination epoch.
In conventional models, the field oscillation during the recombination epoch leads to a cancellation of the birefringence effect and significantly suppresses the EB spectrum of the cosmic microwave background (CMB) polarization.
By introducing an asymmetric potential to the axion, this cancellation becomes incomplete, and a substantial EB spectrum can be produced.
This mechanism also results in a washout of the EE spectrum, which can be probed in future CMB observations.
Our findings suggest the possibility that an axion-like field responsible for isotropic cosmic birefringence can also account for a significant fraction of dark matter.
\end{abstract}

\maketitle

\section{Introduction}

Recently, a hint of a parity-violating signal has been reported in the polarization spectrum of the cosmic microwave background (CMB)~\cite{Minami:2020odp,Diego-Palazuelos:2022dsq,Eskilt:2022wav,Eskilt:2022cff,Eskilt:2023ndm} (see also Ref.~\cite{Komatsu:2022nvu} for a review).
This can be explained by a rotation of the plane of the linear polarization, which is called cosmic birefringence.
The reported cosmic birefringence has an isotropic rotation angle of $\beta \sim 0.3$\,deg~\cite{Minami:2020odp,Diego-Palazuelos:2022dsq,Eskilt:2022wav,Eskilt:2022cff,Eskilt:2023ndm} and has no significant frequency dependence~\cite{Eskilt:2022wav,Eskilt:2022cff}, which is difficult to explain within the framework of the Standard Model of particle physics~\cite{Nakai:2023zdr}.

As the origin of the isotropic cosmic birefringence (ICB), an axion-like particle (ALP) is widely studied~\cite{Fujita:2020ecn,Takahashi:2020tqv,Fung:2021wbz,Nakagawa:2021nme,Jain:2021shf,Choi:2021aze,Obata:2021nql,Nakatsuka:2022epj,Lin:2022niw,Gasparotto:2022uqo,Lee:2022udm,Jain:2022jrp,Murai:2022zur,Gonzalez:2022mcx,Qiu:2023los,Eskilt:2023nxm,Namikawa:2023zux,Gasparotto:2023psh,Ferreira:2023jbu,Greco:2024oie}.
Since the ALP is a parity-odd field coupled to photons, its temporal and spatial variations induce a parity violation in CMB.
When the ALP induces cosmic birefringence, the birefringence angle, $\beta$, is determined only by the ALP-photon coupling and the difference of the ALP field value between the emission and observation.
For appreciable ICB, the ALP field must start to oscillate after the recombination, which requires the ALP mass of $m_\phi < \mathcal{O}(10^{-28})$\,eV.
Otherwise, the birefringence angle oscillates depending on the emission time of CMB photons, and the ICB angle averaged over the line of sight is highly suppressed~\cite{Fujita:2020aqt,Fujita:2020ecn}.

The oscillation of the birefringence angle also washes out the linear polarizations and reduces the TE and EE spectra, although they are parity-even quantities.
This washout effect provides an upper bound on the ALP-photon coupling for ALP dark matter~\cite{Finelli:2008jv,Fedderke:2019ajk}.
Moreover, the ALP oscillation at the observer causes an oscillation in the polarization angle across the entire sky~\cite{Fedderke:2019ajk}.
Recently, a hint of such an oscillating feature in linear polarization has been reported at the significance level of $2.5\sigma$ by the observation of the Crab Nebula~\cite{POLARBEAR:2024vel}. 
Although it is currently in tension with previous constraints~\cite{Fedderke:2019ajk,Reynes:2021bpe,SPT-3G:2022ods,Irsic:2017yje,DES:2020fxi} when interpreted as a signal of ALP dark matter, it motivates us to explore the possibility of the oscillation of the polarization angle.

In this paper, we propose a new mechanism to cause ICB by an ALP that rapidly oscillates during the recombination epoch.
We consider an ALP potential asymmetric under $\phi \leftrightarrow -\phi$.
Then, the field oscillation is also distorted, and the cancellation of the EB spectrum becomes incomplete.
On the other hand, the oscillating birefringence and washout of the TE and EE spectra arise in a similar way to the conventional scenario.
To validate this mechanism, we numerically evaluate the CMB polarization spectra including cosmic birefringence.
We find that the ICB of $\beta \sim 0.3$\,deg can be explained without violating the constraint from the washout effect.
We also provide approximate formulae to obtain the polarization spectra from the spectra without cosmic birefringence.

\section{Cosmic Birefringence}

In general, an axion-like field, $\phi$, is coupled with photons in a parity-violating manner.
We consider a Lagrangian given by
\begin{align}
    \mathcal{L}
    =
    - \frac{1}{2} (\partial_\mu \phi)^2 
    - V(\phi)
    - \frac{1}{4} F_{\mu \nu} F^{\mu \nu}
    - \frac{1}{4} g \phi F_{\mu \nu} \tilde{F}^{\mu \nu}
    \ ,
\end{align}
where $V(\phi)$ is the potential for $\phi$, $F_{\mu \nu}$ is the field strength of the photon field, $\tilde{F}^{\mu \nu}$ is its dual, and $g$ is the ALP-photon coupling constant with mass dimension $-1$.

Under temporal and spatial variations of $\phi$, the polarization plane of linear polarization of photons rotates by an angle, $\beta$, given by~\cite{Carroll:1989vb,Carroll:1991zs,Harari:1992ea}
\begin{align}
    \beta
    = 
    \frac{g}{2}
    \left[ 
        \phi(t_\mathrm{obs}, \bm{x}_\mathrm{obs}) 
        -
        \phi(t_\mathrm{emit}, \bm{x}_\mathrm{emit}) 
    \right]
    \ ,
\end{align}
where $(t_\mathrm{obs}, \bm{x}_\mathrm{obs})$ and $(t_\mathrm{emit}, \bm{x}_\mathrm{emit})$ are the spacetime coordinates at the observation and emission of photons, respectively.
While the spatial dependence of $\phi(t_\mathrm{emit}, \bm{x}_\mathrm{emit})$ induces anisotropic birefringence, the difference between $\phi(t_\mathrm{obs}, \bm{x}_\mathrm{obs})$ and the spatial average of $\phi(t_\mathrm{emit}, \bm{x}_\mathrm{emit})$ induces ICB.
In the following, we focus on the homogeneous component of $\phi(t_\mathrm{emit}, \bm{x}_\mathrm{emit})$ to obtain the isotropic component of the birefringence angle.
Moreover, we are interested in the ALP that starts to oscillate before the recombination, and thus we expect $|\phi(t_\mathrm{obs}, \bm{x}_\mathrm{obs})|$ to be negligibly small.%
\footnote{Throughout this work, we assume that $\phi$ oscillates around the potential minimum at $\phi = 0$.}
Then, the birefringence angle depends only on $\phi$ at the emission and is given by
\begin{align}
    \beta(t_\mathrm{emit})
    \simeq 
    - \frac{g}{2}
    \bar{\phi}(t_\mathrm{emit}) 
    \ ,
\end{align}
where we decompose $\phi$ into its background and fluctuations as $\phi(t, \bm{x}) = \bar{\phi}(t) + \delta \phi(t, \bm{x})$.

To describe the effects of cosmic birefringence on the CMB polarization, we consider the Stokes parameters of linear polarization, $Q \pm iU$, and their Fourier transform, $_{\pm2}\Delta_P(\eta, q, \mu)$.
Here, $\eta$ is the conformal time, $q$ is a wavenumber vector of the Fourier mode, and $\mu \equiv \bm{q}\cdot \bm{k}/(qk)$ is a parameter of the angle between $\bm{q}$ and the photon wavenumber $\bm{k}$.
The Boltzmann equation of $_{\pm2}\Delta_P(\eta, q, \mu)$ is given by~\cite{Liu:2006uh,Finelli:2008jv,Gubitosi:2014cua,Lee:2016jym}
\begin{align}
    &_{\pm 2} \Delta_P' + i q \mu {}_{\pm 2} \Delta_P
    \nonumber \\
    &=
    \tau' \left[ 
        - _{\pm 2} \Delta_P 
        + \sqrt{\frac{6 \pi}{5}} {}_{\pm2} Y_2^0 (\mu) \Pi(\eta, q)
    \right]
    \pm 2 i \beta'  {}_{\pm 2} \Delta_P
    \, ,
\end{align}
where the prime denotes a derivative with respect to $\eta$, $_{\pm2}Y_l^m$ is the spin-2 spherical harmonics, $\Pi$ is the polarization source term~\cite{Zaldarriaga:1996xe}, and $\tau' \equiv a(\eta)n_e(\eta)\sigma_T$ is the differential optical depth with the scale factor $a$, the electron number density $n_e$, and the Thomson scattering cross section $\sigma_T$.
To solve the Boltzmann equation, it is convenient to expand $_{\pm 2} \Delta_P$ with $_{\pm2}Y_l^0$ as
\begin{align}
    _{\pm 2} \Delta_P(\eta,q,\mu)
    \equiv 
    \sum_l i^{-l} \sqrt{4\pi(2l+1)} _{\pm 2} \Delta_{P,l}(\eta, q) {}_{\pm2}Y_l^0(\mu)
    \ .
\end{align}
Then, we can formally integrate the Boltzmann equation as 
\begin{align}
    &_{\pm 2} \Delta_{P, l} (\eta_0, q)
    \nonumber \\
    &=
    -\frac{3}{4} \sqrt{\frac{(l + 2)!}{(l - 2)!}}
    \int_0^{\eta_0} \mathrm{d} \eta \, 
    \tau' e^{-\tau(\eta)} \Pi(\eta, q)
    \frac{j_l(x)}{x^2} e^{\pm 2 i \beta(\eta)}
    \nonumber \\
    &=
    \int_0^{\eta_0} \mathrm{d} \eta \, 
    F(\eta, q) e^{\pm 2 i \beta(\eta)}
    \ ,
    \label{eq: DeltaP integrated Boltzmann}
\end{align}
where $\eta_0$ is the current conformal time, $\tau(\eta) = \int_\eta^{\eta_0} \mathrm{d} \eta_1 \, \tau'(\eta_1)$, $j_l$ is the spherical Bessel function, and $x\equiv q(\eta_0 - \eta)$.
We simplified the factors other than the birefringence effect by $F$ for simplicity of notation.

We further decompose $_{\pm 2} \Delta_{P, l}$ into parity eigenstates, $E$ and $B$ modes, given by~\cite{Zaldarriaga:1996xe, Kamionkowski:1996ks}
\begin{align}
    \Delta_{E,l} \pm \Delta_{B,l}
    \equiv 
    - {}_{\pm 2} \Delta_{P, l} (\eta_0, q)
    \ .
\end{align}
Then, we obtain the polarization angular power spectra as
\begin{align}
    C_l^{XY}
    \equiv 
    4\pi \int \mathrm{d}(\ln q)
    \mathcal{P}_\mathcal{R}(q) 
    \Delta_{X,l}(q) \Delta_{Y,l}(q)
    \ ,
\end{align}
where $X,Y = E$ or $B$, and $\mathcal{P}_\mathcal{R}$ is the primordial curvature power spectrum.

\subsection{Isotropic Cosmic Birefringence}

First, we discuss ICB induced by the time evolution of $\bar{\phi}$.
If $\bar{\phi}$ starts to evolve after the last scattering of the CMB photons, all photons experience the same birefringence angle, $
\beta \simeq g(\bar{\phi}(t_0) - \bar{\phi}_\mathrm{in})/2$, where $\bar{\phi}_\mathrm{in}$ is the initial value of $\bar{\phi}$.
Then, we can factor out $e^{\pm2i\beta}$ in Eq.~\eqref{eq: DeltaP integrated Boltzmann} and obtain
\begin{align}
    \Delta_{E,l} \pm \Delta_{B,l}
    =
    e^{\pm2i \beta} 
    \left( \Delta_{E,l,0} \pm \Delta_{B,l,0} \right)
    \ ,
\end{align}
where the subscript ``0'' denotes quantities without cosmic birefringence.
Then, the polarization angular power spectra under cosmic birefringence are given by~\cite{Lue:1998mq,Feng:2004mq,Liu:2006uh}
\begin{align}
    C_l^{EE}
    &=
    C_{l,0}^{EE} \cos^2(2\beta) + C_{l,0}^{BB} \sin^2(2\beta)
    \ , 
    \\
    C_l^{BB}
    &=
    C_{l,0}^{EE} \sin^2(2\beta) + C_{l,0}^{BB} \cos^2(2\beta)
    \ , 
    \\
    C_l^{EB}
    &=
    \frac{1}{2} \left( C_{l,0}^{EE} - C_{l,0}^{BB} \right)
    \sin(4\beta)
    \ ,
\end{align}
where we assumed $C_{l,0}^{EB} = 0$.
From these relations, we obtain $C_l^{EB} \sim 0.01 C_l^{EE} $ for $\beta \sim 0.3$\,deg and $C_{l,0}^{EE} \gg C_{l,0}^{BB}$.
As a result, by taking an appropriate value of $g$, the reported ICB can be explained by the ALP that evolves only after the recombination epoch.
On the other hand, if $\bar{\phi}$ evolves during the recombination epoch, these relations do not hold. In particular, if $\bar{\phi}$ oscillates, positive and negative contributions to the EB spectrum cancel each other.

\subsection{Washout of CMB polarization}

Next, we analytically estimate the washout effect following Ref.~\cite{Fedderke:2019ajk}.
If the ALP rapidly oscillates during the recombination epoch, $\beta(\eta)$ evolves much faster than $F(\eta, q)$.
Then, we can approximate $_{\pm 2} \Delta_{P, l}$ by dividing the integration over $\eta$ into a sum of the integration over one period of the ALP oscillation:
\begin{align}
    _{\pm 2} \Delta_{P, l} (\eta_0, q)
    &=
    \sum_i
    \int_{\eta_i}^{\eta_i + \delta \eta_i}
    \mathrm{d} \eta \, 
    F(\eta, q)
    e^{\pm 2 i \beta(\eta)}
    \nonumber \\
    &\simeq 
    \sum_i
    F(\eta_i, q)
    \int_{\eta_i}^{\eta_i + \delta \eta_i}
    \mathrm{d} \eta \, 
    e^{\pm 2 i \beta(\eta)}
    \ ,
\end{align}
where $i$ is the label of the ALP oscillation, and $\delta \eta_i$ is the oscillation period.
Then, we approximate $\beta$ in each period by
\begin{align}
    \beta
    \simeq 
    \langle\beta\rangle_i \sin\left(
        2\pi\frac{\eta - \eta_i}{\delta \eta_i}
    \right)
    \ ,
\end{align}
where $\langle\beta\rangle_i$ is the amplitude of $\beta$ at $\eta = \eta_i$.
Consequently, we obtain
\begin{align}
    _{\pm 2} \Delta_{P, l} (\eta_0, q)
    &\simeq 
    \sum_i
    F(\eta_i,q)
    \delta \eta_i
    j_0(2 \langle \beta \rangle_i)
    \nonumber \\
    &\simeq 
    \int \mathrm{d} \eta\,
    F(\eta,q)
    j_0(g \langle \phi \rangle(\eta))
    \nonumber \\
    &\sim 
    j_0(g \langle \phi \rangle_*)
    {}_{\pm 2} \Delta_{P, l, 0} (\eta_0, q)
    \ ,
\end{align}
where $\langle \phi \rangle$ is the oscillation amplitude of the ALP field.
In the last line, we made a further approximation by replacing $j_0(g \langle \phi \rangle(\eta))$ with a representative value at $z = z_* = 1090$.
Finally, we obtain the analytical estimate for the polarization spectra as
\begin{align}
    C_{l,\mathrm{ana}}^{TE}
    &=
    j_0(g \langle \phi \rangle_*)
    C_{l,0}^{TE}
    \ ,
    \label{eq: washout formula TE}
    \\
    C_{l,\mathrm{ana}}^{EE}
    &=
    j_0^2(g \langle \phi \rangle_*)
    C_{l,0}^{EE}
    \ ,
    \label{eq: washout formula EE}
\end{align}
which is reduced to 
\begin{align}
    C_{l,\mathrm{ana}}^{TE}
    &\simeq 
    \left(1 - \frac{1}{4}g^2 \langle \phi \rangle_*^2 \right)
    C_{l,0}^{TE}
    \ ,
    \label{eq: washout formula TE approx}
    \\
    C_{l,\mathrm{ana}}^{EE}
    &\simeq 
    \left(1 - \frac{1}{2}g^2 \langle \phi \rangle_*^2 \right)
    C_{l,0}^{EE}
    \ ,
    \label{eq: washout formula EE approx}
\end{align}
in the limit of a small oscillation angle.

\section{Numerical results}

Here, we show the numerical results for the CMB polarization spectra under oscillating birefringence.
Let us consider an asymmetric potential.
Since this study aims to generally investigate the effects of an asymmetric potential, we adopt a simple potential given by
\begin{align}
    V(\phi)
    =
    \frac{m_\phi^2}{2} \phi^2 
    + C_3 \phi^3
    + C_4 \phi^4
    \ ,
    \label{eq: distorted potential}
\end{align}
where $C_3$ and $C_4$ are constants.
\begin{figure}[t]
    \centering
    \includegraphics[width=.45\textwidth]{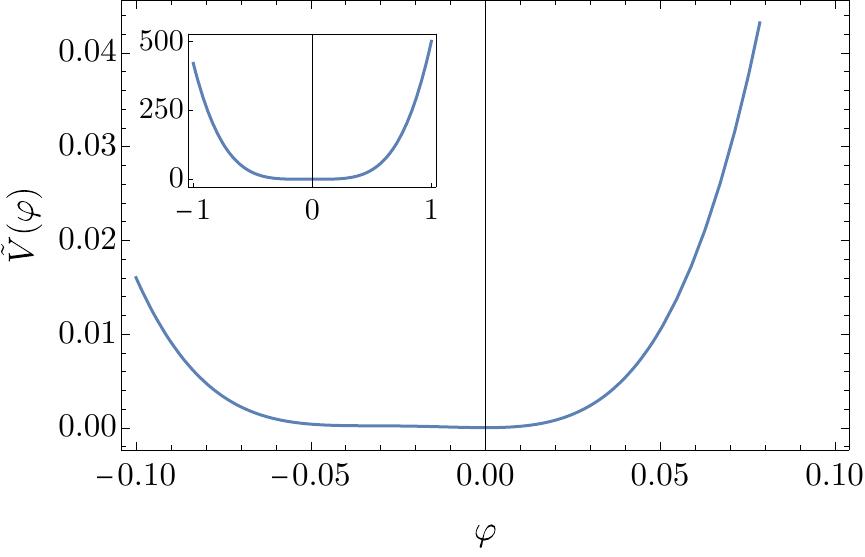}
    \caption{%
        Asymmetric potential $\tilde{V}(\varphi)$ normalized as in Eq.~\eqref{eq: normalized potential}.
        Here we use $c_3 = 40$ and $c_4 = 460$.
    }
    \label{fig: V potential}
    \end{figure}
Such an asymmetric potential can be realized, e.g., in the context of axion monodromy~\cite{McAllister:2008hb}, where the potential is given by the sum of a linear term and a cosine-type potential.
Alternatively, a superposition of cosine potentials with different scales and phases also gives an asymmetric potential around its minima.

In numerical simulations, we normalize the ALP field by $\varphi \equiv \phi/\phi_\mathrm{in}$ with the initial field value $\phi_\mathrm{in}$, and parameterize the asymmetric potential by
\begin{align}
    V(\varphi)
    &=
    \frac{m_\phi^2 \phi_\mathrm{in}^2}{2} \left(
        \varphi^2 + c_3 \varphi^3 + c_4 \varphi^4
    \right)
    \nonumber \\
    &\equiv
    \frac{m_\phi^2 \phi_\mathrm{in}^2}{2} \tilde{V}(\varphi)
    \ ,
    \label{eq: normalized potential}
\end{align}
where $c_3$ and $c_4$ are dimensionless constants.
This potential is approximated by a quadratic potential for $\varphi \ll 1/c_3, 1/\sqrt{c_4}$ and a quartic potential for $\varphi \gg 1/c_3, 1/\sqrt{c_4}$.
In the intermediate region, the shape of the potential can be highly distorted.
In the following, we assume that $V(\varphi)$ has a unique local minimum, which requires $c_4 > 9 c_3^2/32$.
We show the shape of the asymmetric potential for $c_3 = 40$ and $c_4 = 460$ in Fig.~\ref{fig: V potential}.

To evaluate the CMB spectra with cosmic birefringence, we assume that the energy density of the ALP is negligibly small in the epoch of interest.
Then, we can deal with the background cosmology and the ALP dynamics separately.
We first solve the equation of motion for the axion:
\begin{align}
    \ddot{\varphi} + 3 H \dot{\varphi} 
    +m_\phi^2 \frac{\partial \tilde{V}(\varphi)}{\partial \varphi}
    =
    0
    \ ,
\end{align}
where we use the background evolution obtained by the CMB Boltzmann solver \texttt{CLASS}~\cite{Lesgourgues:2011re} for the Hubble parameter $H$.
We translate the solution $\varphi(t)$ into the birefringence angle $\beta(t)$ by fixing $g \phi_\mathrm{in}$.
Then, we obtain the CMB spectra using \texttt{birefCLASS}~\cite{Nakatsuka:2022epj, Naokawa:2023upt} based on \texttt{CLASS}.
In this work, we focus on scalar perturbations and ignore tensor perturbations.

\subsection{Mass potential}

\begin{figure}[t]
    \centering
    \includegraphics[width=.45\textwidth]{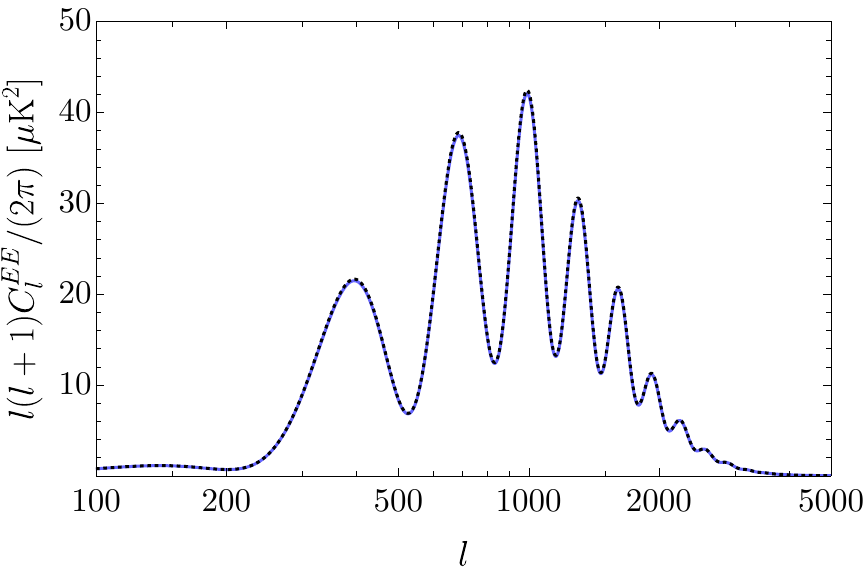}
    \caption{%
        EE spectrum with and without cosmic birefringence induced by the ALP with a mass potential for $m_\phi = 10^{-26}$\,eV and $g \phi_\mathrm{in}/2 = 326$\,deg.
        The blue lines denote the spectra with birefringence, and the black-dashed lines denote the EE spectrum without birefringence.
        The former is slightly suppressed compared with the latter.
    }
    \label{fig: EE}
    \end{figure}

Let us begin with the mass potential case, $c_3 = c_4 = 0$, with $m_\phi = 10^{-26}$\,eV.
We show the EE spectrum in Fig.~\ref{fig: EE}. 
Here, we set $g \phi_\mathrm{in}/2 = 326$\,deg.
For comparison, we also show the EE spectra without birefringence, which is slightly larger than that with birefringence.
The EB spectrum is highly suppressed as $|C_l^{EB}| < 10^{-6} C_l^{EE}$ due to the oscillating $\beta(\eta)$.

Using this result, we can check the validity of the analytical formula for the EE spectrum~\eqref{eq: washout formula EE}.
We evaluate the ALP amplitude at the recombination, $\langle \phi \rangle_*$, by 
\begin{align}
    \langle \phi \rangle_*^2
    =
    \phi_\mathrm{in}^2 \left( 
        \varphi_*^2 + \frac{\dot{\varphi}_*^2}{m_\phi^2}
    \right)
    \ ,
\end{align}
where the subscript $*$ denotes the quantities at $z = z_*$.
From the numerical solution of $\varphi$, we obtain 
$g \langle \phi \rangle_*/2 \simeq 4.0^\circ \simeq 0.07$\,rad.
Then, we can estimate the suppressed EE spectrum as 
\begin{align}
    C_l^{EE}
    \simeq 
    0.99 C_{l,0}^{EE}
    \ ,
\end{align}
which saturates the upper bound from the CMB observations~\cite{Fedderke:2019ajk}.

\begin{figure}[t]
    \centering
    \includegraphics[width=.45\textwidth]{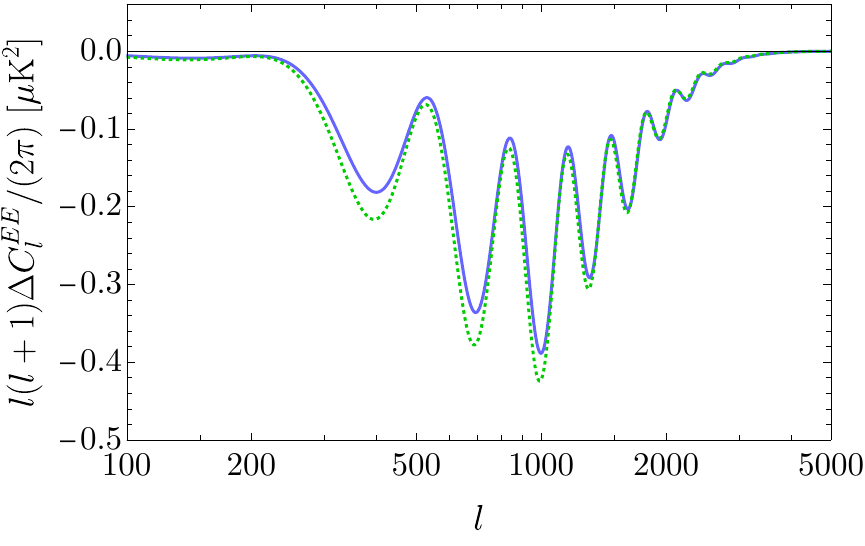}
    \caption{%
        Suppression of the EE spectrum due to the washout effect, $\Delta C_l^{EE} \equiv C_l^{EE} - C_{l,0}^{EE}$.
        The blue and green dashed lines denote the numerical result and analytical estimate, respectively.
    }
    \label{fig: dEE}
    \end{figure}

We show the comparison of the numerical result and analytical estimate in Fig.~\ref{fig: dEE}.
The analytical estimate well reproduces the numerical result, and their difference is at most $10\%$ of $\Delta C_l^{EE}\equiv C_l^{EE} - C_{l,0}^{EE}$.
We expect that this difference comes from the time evolution of $\langle \phi \rangle$, which is not captured in the analytical formula.
The constraints on the washout effect will be improved by a factor of $7$ in terms of $\Delta C_l^{EE}$ for the cosmic-variance limited observations~\cite{Fedderke:2019ajk}.
Thus, the analytical formula is sufficiently accurate at least for near-future CMB observations.

\subsection{Asymmetric potential}

\begin{figure}[t]
    \centering
    \includegraphics[width=.45\textwidth]{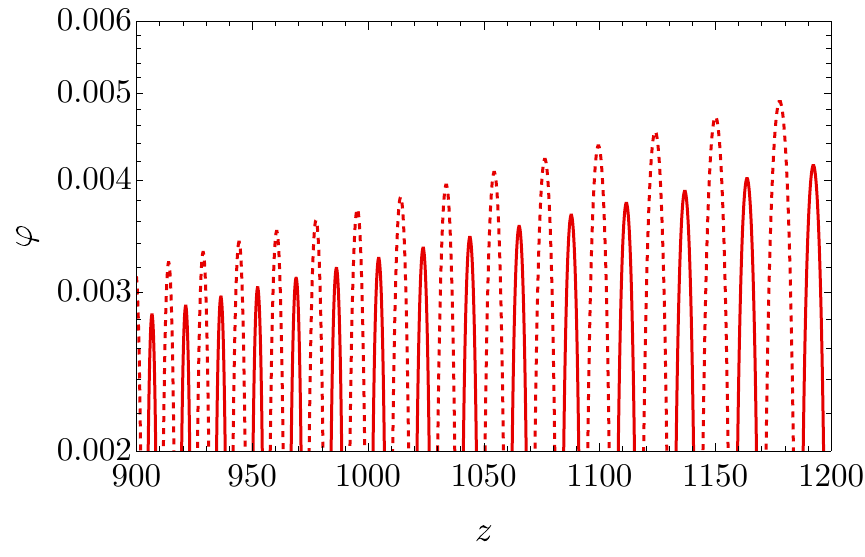}
    \caption{%
        Time evolution of $\varphi$ for $m_\phi = 10^{-26}$\,eV, $c_3 = 40$, and $c_4 = 460$.
        The solid and dashed line shows $\varphi$ and $-\varphi$, respectively.
    }
    \label{fig: phi evolution mod}
    \end{figure}

Next, we consider the asymmetric potential.
Here, we use $m_\phi = 10^{-26}$\,eV, $c_3 = 40$, and $c_4 = 460$.
We show the numerical solution for $\varphi$ in Fig.~\ref{fig: phi evolution mod}.
Due to the distortion in the potential, the oscillation amplitudes for positive and negative $\varphi$ are different.
We chose $c_3$ and $c_4$ so that the distortion of the potential is effective around the recombination epoch.

\begin{figure}[t]
    \centering
    \includegraphics[width=.45\textwidth]{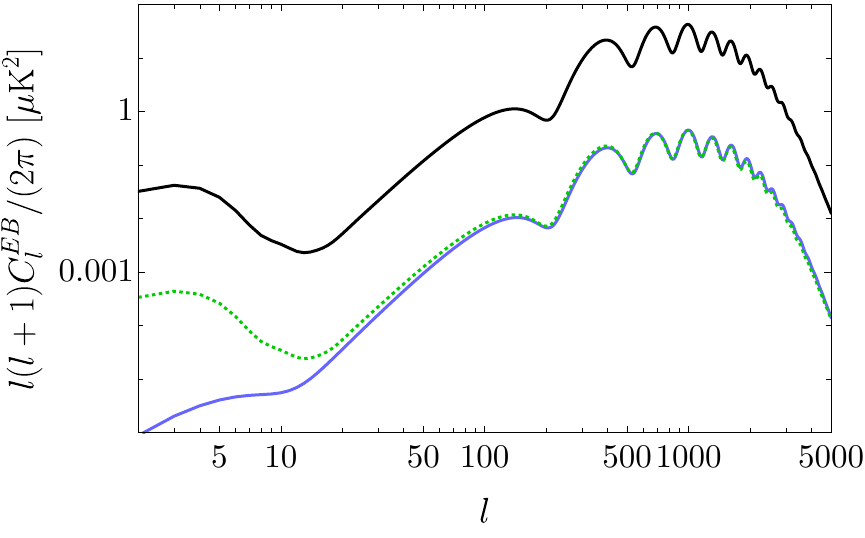}
    \caption{%
        EB spectrum for $m_\phi = 10^{-26}$\,eV, $c_3 = 40$, and $c_4 = 460$.
        The blue and green-dotted lines denote the numerical result and analytical estimate, respectively.
        For comparison, we also show the EE spectrum by the black line.
    }
    \label{fig: EB mod}
    \end{figure}

In this case, the EE spectrum is similar to the previous case, and the EB spectrum also becomes largely proportional to $C_{l,0}^{EE}$ in contrast to the previous case.
We show the numerical result for the EB spectra in Fig.~\ref{fig: EB mod}.
Here, we set $g \phi_\mathrm{in} = 691$\,deg, which results in the EB spectrum corresponding to $\beta = 0.3$\,deg.
The EB spectrum does not have the reionization bump at $l < 10$, which comes from linear polarization produced during the reionization epoch (see Refs.~\cite{Sherwin:2021vgb,Nakatsuka:2022epj}).

We also present analytical estimates for the EE and EB spectra.
Due to the asymmetric shape of the potential, we cannot represent the oscillation of $\varphi$ by a sine function.
Here, we pick up one oscillation period from $t_1$ to $t_2$ around $z = z_*$ by finding the zero points of $\varphi(t)$.
Then, we define an effective oscillation amplitude $\langle \phi \rangle_{*,\mathrm{eff}}$ by
\begin{align}
    \langle \phi \rangle_{*,\mathrm{eff}}
    &\equiv 
    \frac{\pi \phi_\mathrm{in}}{2(t_2-t_1)} \int_{t_1}^{t_2} \mathrm{d} t\,
    |\varphi(t)|
    \ .
    \label{eq: phi amplitude mod}
\end{align}
This definition reproduces the oscillation amplitude when $\varphi$ oscillates following a sine function.
We also obtain the averaged field value $\bar{\phi}_*$ during the recombination epoch weighed by the visibility function as 
\begin{align}
    \bar{\phi}_*
    &\equiv 
    \phi_\mathrm{in}
    \frac{\int_{\eta_\mathrm{min}}^{\eta_\mathrm{max}} \mathrm{d} \eta\,
    \tau' e^{-\tau(\eta)} \varphi(\eta)}
    {\int_{\eta_\mathrm{min}}^{\eta_\mathrm{max}} \mathrm{d} \eta\,
    \tau' e^{-\tau(\eta)} }
    \ ,
    \label{eq: phi average mod}
\end{align}
where we set $\eta_\mathrm{min}$ and $\eta_\mathrm{max}$ to the conformal time at $z = 10^5$ and $100$, respectively.
Since the asymmetry of the $\varphi$ oscillation evolves in time as well as the amplitude, we adopted the average over a long period for $\bar{\phi}_*$ rather than the average over one period to maintain the accuracy of the approximation.
Using these quantities, we estimate the spectra as
\begin{align}
    C_{l,\mathrm{ana}}^{EE}
    =
    j_0^2(g \langle \phi \rangle_{*,\mathrm{eff}}) C_{l,0}^{EE}
    \simeq 
    0.995 C_{l,0}^{EE}
    \ ,
    \\
    C_{l,\mathrm{ana}}^{EB}
    =
    \frac{1}{2} C_{l,0}^{EE} \sin(2 g \bar{\phi}_*)
    \simeq 
    0.010 C_{l,0}^{EE}
    \ ,
\end{align}
from the numerical solution of $\varphi$.
While the washout effect of the EE spectrum is smaller than the current constraint, the EB spectrum reproduces $\beta \sim 0.3$\,deg, which gives $\sin(4\beta)/2 \simeq 0.010$.
For $l > 50$, $C_{l,\mathrm{ana}}^{EB}$ reproduces the numerical result with an accuracy of 25\%.

In this example, we have obtained the EB spectrum roughly corresponding to $\beta \sim 0.3$\,deg without violating the constraint on the washout effect.
Since $\langle \phi \rangle_{*,\mathrm{eff}}$ and $\bar{\phi}_*$ are determined by the potential in different ways, we can obtain larger or smaller EB spectra with a fixed magnitude of the washout effect by tuning the potential parameters.

Finally, we comment on the dependence of the EB spectrum on the model parameters.
To obtain a substantial EB signal, the asymmetric oscillation during the recombination epoch is essential (see Fig.~\ref{fig: phi evolution mod}).
To realize such a situation, the quadratic, cubic, and quartic terms should contribute comparably to the potential at that time, which is not the case for general choices of $m_\phi$, $c_3$, and $c_4$.
If there are higher-order terms in the potential, they should not dominate over the lower-order terms during the recombination epoch to realize the asymmetric oscillation.
Note that the dependence on $\phi_\mathrm{in}$ is absorbed to $c_3$ and $c_4$ for fixed $m_\phi$ in our notation of the potential~\eqref{eq: normalized potential}.

\section{Summary}

In this paper, we investigated the effect of oscillating birefringence on the CMB polarization spectra.
First, we numerically evaluated the EE and EB spectra for the mass potential and validated the analytical estimate.
Then, we investigated an asymmetric potential and found that the oscillating birefringence by an asymmetric potential can explain the ICB with $\beta \sim 0.3$\,deg without violating the constraint from the washout effect.
We also provided analytical formulae for the EE and EB spectra with an asymmetric potential.

Although we used $m_\phi = 10^{-26}$\,eV as an example, we have checked that the result is almost independent of $m_\phi$ as long as $\phi$ rapidly oscillates during the recombination epoch.
Rather, the resultant EE and EB spectra are largely determined by $\langle \phi \rangle_{*,\mathrm{eff}}$ and $\bar{\phi}_*$ given in Eqs.~\eqref{eq: phi amplitude mod} and \eqref{eq: phi average mod}.
Thus, $\phi$ will be able to explain the ICB with $\beta \sim 0.3$\,deg even for $m_\phi \gg 10^{-26}$\,eV.
This opens up the possibility that $\phi$ is a significant fraction of dark matter and the origin of the ICB at the same time. 
Considering the X-ray constraint of $g < 6.3 \times 10^{-13}\,\mathrm{GeV}^{-1}$ for the ultra-light mass region~\cite{Reynes:2021bpe}, we require $\bar{\phi}_* \gtrsim 10^{10}$\,GeV to obtain $\beta \sim 0.3$\,deg.
Since the energy density of the ALP is at most that of dark matter, we require that $m_\phi^2 \bar{\phi}_*^2/2$ should be smaller than the dark matter density at the recombination as a rough necessary condition.
Consequently, we obtain the upper bound of $m_\phi \lesssim 10^{-20}$\,eV.
This constraint becomes more severe when $\bar{\phi}_*$ is suppressed compared to $\langle \phi \rangle_{*,\mathrm{eff}}$.

In this work, we assumed that the ALP is a subdominant component in the universe and adopted a toy-model potential.
To test the possibility of the ICB by ALP dark matter, we need to solve the background evolution including the ALP and the CMB spectra consistently.
Since the time scale of the ALP oscillation is much shorter than that of the CMB physics, it will be computationally expensive.
However, an approximate analytical solution for the evolution of $\phi(t)$ could be used to speed up the calculations~\cite{Galaverni:2023zhv}.
Moreover, the ALP potential should be modified for dark matter because the ALP does not behave as non-relativistic matter at the beginning of oscillations with the quartic potential.
For instance, the combination of cosine-type potentials can lead to a potential distorted only in a certain region between the initial value and the potential minimum.
Further detailed investigation of this possibility is left for future work.

\begin{acknowledgments}
We are grateful to Kohei Kamada, Eiichiro Komatsu, Toshiya Namikawa, Fumihiro Naokawa, Ippei Obata, Maresuke Shiraishi, and Wen Yin for helpful discussions and comments.
This work was supported in part by JSPS KAKENHI Grant Numbers 20H05859, 23KJ0088, and 24K17039.
\end{acknowledgments}

\bibliographystyle{JHEP}
\bibliography{Ref}

\providecommand{\href}[2]{#2}\begingroup\raggedright\begin{thebibliography}{10}

\bibitem{Minami:2020odp}
Y.~Minami and E.~Komatsu, \emph{{New Extraction of the Cosmic Birefringence
  from the Planck 2018 Polarization Data}},
  \href{https://doi.org/10.1103/PhysRevLett.125.221301}{\emph{Phys. Rev. Lett.}
  {\bfseries 125} (2020) 221301},
  [\href{https://arxiv.org/abs/2011.11254}{{\ttfamily 2011.11254}}].

\bibitem{Diego-Palazuelos:2022dsq}
P.~Diego-Palazuelos et~al., \emph{{Cosmic Birefringence from the Planck Data
  Release 4}},
  \href{https://doi.org/10.1103/PhysRevLett.128.091302}{\emph{Phys. Rev. Lett.}
  {\bfseries 128} (2022) 091302},
  [\href{https://arxiv.org/abs/2201.07682}{{\ttfamily 2201.07682}}].

\bibitem{Eskilt:2022wav}
J.~R. Eskilt, \emph{{Frequency-dependent constraints on cosmic birefringence
  from the LFI and HFI Planck Data Release 4}},
  \href{https://doi.org/10.1051/0004-6361/202243269}{\emph{Astron. Astrophys.}
  {\bfseries 662} (2022) A10},
  [\href{https://arxiv.org/abs/2201.13347}{{\ttfamily 2201.13347}}].

\bibitem{Eskilt:2022cff}
J.~R. Eskilt and E.~Komatsu, \emph{{Improved constraints on cosmic
  birefringence from the WMAP and Planck cosmic microwave background
  polarization data}},
  \href{https://doi.org/10.1103/PhysRevD.106.063503}{\emph{Phys. Rev. D}
  {\bfseries 106} (2022) 063503},
  [\href{https://arxiv.org/abs/2205.13962}{{\ttfamily 2205.13962}}].

\bibitem{Eskilt:2023ndm}
{\scshape Cosmoglobe} collaboration, J.~R. Eskilt et~al., \emph{{COSMOGLOBE DR1
  results - II. Constraints on isotropic cosmic birefringence from reprocessed
  WMAP and Planck LFI data}},
  \href{https://doi.org/10.1051/0004-6361/202346829}{\emph{Astron. Astrophys.}
  {\bfseries 679} (2023) A144},
  [\href{https://arxiv.org/abs/2305.02268}{{\ttfamily 2305.02268}}].

\bibitem{Komatsu:2022nvu}
E.~Komatsu, \emph{{New physics from the polarized light of the cosmic microwave
  background}}, \href{https://doi.org/10.1038/s42254-022-00452-4}{\emph{Nature
  Rev. Phys.} {\bfseries 4} (2022) 452--469},
  [\href{https://arxiv.org/abs/2202.13919}{{\ttfamily 2202.13919}}].

\bibitem{Nakai:2023zdr}
Y.~Nakai, R.~Namba, I.~Obata, Y.-C. Qiu and R.~Saito, \emph{{Can we explain
  cosmic birefringence without a new light field beyond Standard Model?}},
  \href{https://doi.org/10.1007/JHEP01(2024)057}{\emph{JHEP} {\bfseries 01}
  (2024) 057}, [\href{https://arxiv.org/abs/2310.09152}{{\ttfamily
  2310.09152}}].

\bibitem{Fujita:2020ecn}
T.~Fujita, K.~Murai, H.~Nakatsuka and S.~Tsujikawa, \emph{{Detection of
  isotropic cosmic birefringence and its implications for axionlike particles
  including dark energy}},
  \href{https://doi.org/10.1103/PhysRevD.103.043509}{\emph{Phys. Rev. D}
  {\bfseries 103} (2021) 043509},
  [\href{https://arxiv.org/abs/2011.11894}{{\ttfamily 2011.11894}}].

\bibitem{Takahashi:2020tqv}
F.~Takahashi and W.~Yin, \emph{{Kilobyte Cosmic Birefringence from ALP Domain
  Walls}}, \href{https://doi.org/10.1088/1475-7516/2021/04/007}{\emph{JCAP}
  {\bfseries 04} (2021) 007},
  [\href{https://arxiv.org/abs/2012.11576}{{\ttfamily 2012.11576}}].

\bibitem{Fung:2021wbz}
L.~W.~H. Fung, L.~Li, T.~Liu, H.~N. Luu, Y.-C. Qiu and S.~H.~H. Tye,
  \emph{{Axi-Higgs cosmology}},
  \href{https://doi.org/10.1088/1475-7516/2021/08/057}{\emph{JCAP} {\bfseries
  08} (2021) 057}, [\href{https://arxiv.org/abs/2102.11257}{{\ttfamily
  2102.11257}}].

\bibitem{Nakagawa:2021nme}
S.~Nakagawa, F.~Takahashi and M.~Yamada, \emph{{Cosmic Birefringence Triggered
  by Dark Matter Domination}},
  \href{https://doi.org/10.1103/PhysRevLett.127.181103}{\emph{Phys. Rev. Lett.}
  {\bfseries 127} (2021) 181103},
  [\href{https://arxiv.org/abs/2103.08153}{{\ttfamily 2103.08153}}].

\bibitem{Jain:2021shf}
M.~Jain, A.~J. Long and M.~A. Amin, \emph{{CMB birefringence from
  ultralight-axion string networks}},
  \href{https://doi.org/10.1088/1475-7516/2021/05/055}{\emph{JCAP} {\bfseries
  05} (2021) 055}, [\href{https://arxiv.org/abs/2103.10962}{{\ttfamily
  2103.10962}}].

\bibitem{Choi:2021aze}
G.~Choi, W.~Lin, L.~Visinelli and T.~T. Yanagida, \emph{{Cosmic birefringence
  and electroweak axion dark energy}},
  \href{https://doi.org/10.1103/PhysRevD.104.L101302}{\emph{Phys. Rev. D}
  {\bfseries 104} (2021) L101302},
  [\href{https://arxiv.org/abs/2106.12602}{{\ttfamily 2106.12602}}].

\bibitem{Obata:2021nql}
I.~Obata, \emph{{Implications of the cosmic birefringence measurement for the
  axion dark matter search}},
  \href{https://doi.org/10.1088/1475-7516/2022/09/062}{\emph{JCAP} {\bfseries
  09} (2022) 062}, [\href{https://arxiv.org/abs/2108.02150}{{\ttfamily
  2108.02150}}].

\bibitem{Nakatsuka:2022epj}
H.~Nakatsuka, T.~Namikawa and E.~Komatsu, \emph{{Is cosmic birefringence due to
  dark energy or dark matter? A tomographic approach}},
  \href{https://doi.org/10.1103/PhysRevD.105.123509}{\emph{Phys. Rev. D}
  {\bfseries 105} (2022) 123509},
  [\href{https://arxiv.org/abs/2203.08560}{{\ttfamily 2203.08560}}].

\bibitem{Lin:2022niw}
W.~Lin and T.~T. Yanagida, \emph{{Consistency of the string inspired
  electroweak axion with cosmic birefringence}},
  \href{https://doi.org/10.1103/PhysRevD.107.L021302}{\emph{Phys. Rev. D}
  {\bfseries 107} (2023) L021302},
  [\href{https://arxiv.org/abs/2208.06843}{{\ttfamily 2208.06843}}].

\bibitem{Gasparotto:2022uqo}
S.~Gasparotto and I.~Obata, \emph{{Cosmic birefringence from monodromic axion
  dark energy}},
  \href{https://doi.org/10.1088/1475-7516/2022/08/025}{\emph{JCAP} {\bfseries
  08} (2022) 025}, [\href{https://arxiv.org/abs/2203.09409}{{\ttfamily
  2203.09409}}].

\bibitem{Lee:2022udm}
N.~Lee, S.~C. Hotinli and M.~Kamionkowski, \emph{{Probing cosmic birefringence
  with polarized Sunyaev-Zel\textquoteright{}dovich tomography}},
  \href{https://doi.org/10.1103/PhysRevD.106.083518}{\emph{Phys. Rev. D}
  {\bfseries 106} (2022) 083518},
  [\href{https://arxiv.org/abs/2207.05687}{{\ttfamily 2207.05687}}].

\bibitem{Jain:2022jrp}
M.~Jain, R.~Hagimoto, A.~J. Long and M.~A. Amin, \emph{{Searching for
  axion-like particles through CMB birefringence from string-wall networks}},
  \href{https://doi.org/10.1088/1475-7516/2022/10/090}{\emph{JCAP} {\bfseries
  10} (2022) 090}, [\href{https://arxiv.org/abs/2208.08391}{{\ttfamily
  2208.08391}}].

\bibitem{Murai:2022zur}
K.~Murai, F.~Naokawa, T.~Namikawa and E.~Komatsu, \emph{{Isotropic cosmic
  birefringence from early dark energy}},
  \href{https://doi.org/10.1103/PhysRevD.107.L041302}{\emph{Phys. Rev. D}
  {\bfseries 107} (2023) L041302},
  [\href{https://arxiv.org/abs/2209.07804}{{\ttfamily 2209.07804}}].

\bibitem{Gonzalez:2022mcx}
D.~Gonzalez, N.~Kitajima, F.~Takahashi and W.~Yin, \emph{{Stability of domain
  wall network with initial inflationary fluctuations and its implications for
  cosmic birefringence}},
  \href{https://doi.org/10.1016/j.physletb.2023.137990}{\emph{Phys. Lett. B}
  {\bfseries 843} (2023) 137990},
  [\href{https://arxiv.org/abs/2211.06849}{{\ttfamily 2211.06849}}].

\bibitem{Qiu:2023los}
Y.-C. Qiu, J.-W. Wang and T.~T. Yanagida, \emph{{High-Quality Axions in a Class
  of Chiral U(1) Gauge Theories}},
  \href{https://doi.org/10.1103/PhysRevLett.131.071802}{\emph{Phys. Rev. Lett.}
  {\bfseries 131} (2023) 071802},
  [\href{https://arxiv.org/abs/2301.02345}{{\ttfamily 2301.02345}}].

\bibitem{Eskilt:2023nxm}
J.~R. Eskilt, L.~Herold, E.~Komatsu, K.~Murai, T.~Namikawa and F.~Naokawa,
  \emph{{Constraints on Early Dark Energy from Isotropic Cosmic
  Birefringence}},
  \href{https://doi.org/10.1103/PhysRevLett.131.121001}{\emph{Phys. Rev. Lett.}
  {\bfseries 131} (2023) 121001},
  [\href{https://arxiv.org/abs/2303.15369}{{\ttfamily 2303.15369}}].

\bibitem{Namikawa:2023zux}
T.~Namikawa and I.~Obata, \emph{{Cosmic birefringence tomography with polarized
  Sunyaev-Zel\textquoteright{}dovich effect}},
  \href{https://doi.org/10.1103/PhysRevD.108.083510}{\emph{Phys. Rev. D}
  {\bfseries 108} (2023) 083510},
  [\href{https://arxiv.org/abs/2306.08875}{{\ttfamily 2306.08875}}].

\bibitem{Gasparotto:2023psh}
S.~Gasparotto and E.~I. Sfakianakis, \emph{{Cosmic birefringence from the
  Axiverse}}, \href{https://doi.org/10.1088/1475-7516/2023/11/017}{\emph{JCAP}
  {\bfseries 11} (2023) 017},
  [\href{https://arxiv.org/abs/2306.16355}{{\ttfamily 2306.16355}}].

\bibitem{Ferreira:2023jbu}
R.~Z. Ferreira, S.~Gasparotto, T.~Hiramatsu, I.~Obata and O.~Pujolas,
  \emph{{Axionic defects in the CMB: birefringence and gravitational waves}},
  \href{https://doi.org/10.1088/1475-7516/2024/05/066}{\emph{JCAP} {\bfseries
  05} (2024) 066}, [\href{https://arxiv.org/abs/2312.14104}{{\ttfamily
  2312.14104}}].

\bibitem{Greco:2024oie}
A.~Greco, N.~Bartolo and A.~Gruppuso, \emph{{A New Solution for the Observed
  Isotropic Cosmic Birefringence Angle and its Implications for the Anisotropic
  Counterpart through a Boltzmann Approach}},
  \href{https://arxiv.org/abs/2401.07079}{{\ttfamily 2401.07079}}.

\bibitem{Fujita:2020aqt}
T.~Fujita, Y.~Minami, K.~Murai and H.~Nakatsuka, \emph{{Probing axionlike
  particles via cosmic microwave background polarization}},
  \href{https://doi.org/10.1103/PhysRevD.103.063508}{\emph{Phys. Rev. D}
  {\bfseries 103} (2021) 063508},
  [\href{https://arxiv.org/abs/2008.02473}{{\ttfamily 2008.02473}}].

\bibitem{Finelli:2008jv}
F.~Finelli and M.~Galaverni, \emph{{Rotation of Linear Polarization Plane and
  Circular Polarization from Cosmological Pseudo-Scalar Fields}},
  \href{https://doi.org/10.1103/PhysRevD.79.063002}{\emph{Phys. Rev. D}
  {\bfseries 79} (2009) 063002},
  [\href{https://arxiv.org/abs/0802.4210}{{\ttfamily 0802.4210}}].

\bibitem{Fedderke:2019ajk}
M.~A. Fedderke, P.~W. Graham and S.~Rajendran, \emph{{Axion Dark Matter
  Detection with CMB Polarization}},
  \href{https://doi.org/10.1103/PhysRevD.100.015040}{\emph{Phys. Rev. D}
  {\bfseries 100} (2019) 015040},
  [\href{https://arxiv.org/abs/1903.02666}{{\ttfamily 1903.02666}}].

\bibitem{POLARBEAR:2024vel}
{\scshape POLARBEAR} collaboration, S.~Adachi et~al., \emph{{Exploration of the
  polarization angle variability of the Crab Nebula with POLARBEAR and its
  application to the search for axion-like particles}},
  \href{https://arxiv.org/abs/2403.02096}{{\ttfamily 2403.02096}}.

\bibitem{Reynes:2021bpe}
J.~S. Reyn\'es, J.~H. Matthews, C.~S. Reynolds, H.~R. Russell, R.~N. Smith and
  M.~C.~D. Marsh, \emph{{New constraints on light axion-like particles using
  Chandra transmission grating spectroscopy of the powerful cluster-hosted
  quasar H1821+643}}, \href{https://doi.org/10.1093/mnras/stab3464}{\emph{Mon.
  Not. Roy. Astron. Soc.} {\bfseries 510} (2021) 1264--1277},
  [\href{https://arxiv.org/abs/2109.03261}{{\ttfamily 2109.03261}}].

\bibitem{SPT-3G:2022ods}
{\scshape SPT-3G} collaboration, K.~R. Ferguson et~al., \emph{{Searching for
  axionlike time-dependent cosmic birefringence with data from SPT-3G}},
  \href{https://doi.org/10.1103/PhysRevD.106.042011}{\emph{Phys. Rev. D}
  {\bfseries 106} (2022) 042011},
  [\href{https://arxiv.org/abs/2203.16567}{{\ttfamily 2203.16567}}].

\bibitem{Irsic:2017yje}
V.~Ir\v{s}i\v{c}, M.~Viel, M.~G. Haehnelt, J.~S. Bolton and G.~D. Becker,
  \emph{{First constraints on fuzzy dark matter from Lyman-$\alpha$ forest data
  and hydrodynamical simulations}},
  \href{https://doi.org/10.1103/PhysRevLett.119.031302}{\emph{Phys. Rev. Lett.}
  {\bfseries 119} (2017) 031302},
  [\href{https://arxiv.org/abs/1703.04683}{{\ttfamily 1703.04683}}].

\bibitem{DES:2020fxi}
{\scshape DES} collaboration, E.~O. Nadler et~al., \emph{{Milky Way Satellite
  Census. III. Constraints on Dark Matter Properties from Observations of Milky
  Way Satellite Galaxies}},
  \href{https://doi.org/10.1103/PhysRevLett.126.091101}{\emph{Phys. Rev. Lett.}
  {\bfseries 126} (2021) 091101},
  [\href{https://arxiv.org/abs/2008.00022}{{\ttfamily 2008.00022}}].

\bibitem{Carroll:1989vb}
S.~M. Carroll, G.~B. Field and R.~Jackiw, \emph{{Limits on a Lorentz and Parity
  Violating Modification of Electrodynamics}},
  \href{https://doi.org/10.1103/PhysRevD.41.1231}{\emph{Phys. Rev. D}
  {\bfseries 41} (1990) 1231}.

\bibitem{Carroll:1991zs}
S.~M. Carroll and G.~B. Field, \emph{{The Einstein equivalence principle and
  the polarization of radio galaxies}},
  \href{https://doi.org/10.1103/PhysRevD.43.3789}{\emph{Phys. Rev. D}
  {\bfseries 43} (1991) 3789}.

\bibitem{Harari:1992ea}
D.~Harari and P.~Sikivie, \emph{{Effects of a Nambu-Goldstone boson on the
  polarization of radio galaxies and the cosmic microwave background}},
  \href{https://doi.org/10.1016/0370-2693(92)91363-E}{\emph{Phys. Lett. B}
  {\bfseries 289} (1992) 67--72}.

\bibitem{Liu:2006uh}
G.-C. Liu, S.~Lee and K.-W. Ng, \emph{{Effect on cosmic microwave background
  polarization of coupling of quintessence to pseudoscalar formed from the
  electromagnetic field and its dual}},
  \href{https://doi.org/10.1103/PhysRevLett.97.161303}{\emph{Phys. Rev. Lett.}
  {\bfseries 97} (2006) 161303},
  [\href{https://arxiv.org/abs/astro-ph/0606248}{{\ttfamily
  astro-ph/0606248}}].

\bibitem{Gubitosi:2014cua}
G.~Gubitosi, M.~Martinelli and L.~Pagano, \emph{{Including birefringence into
  time evolution of CMB: current and future constraints}},
  \href{https://doi.org/10.1088/1475-7516/2014/12/020}{\emph{JCAP} {\bfseries
  12} (2014) 020}, [\href{https://arxiv.org/abs/1410.1799}{{\ttfamily
  1410.1799}}].

\bibitem{Lee:2016jym}
S.~Lee, G.-C. Liu and K.-W. Ng, \emph{{Dark Ultra-Light Scalars and Cosmic
  Parity Violation}}, {\emph{The Universe} {\bfseries 4} (2016) 29--44},
  [\href{https://arxiv.org/abs/1912.12903}{{\ttfamily 1912.12903}}].

\bibitem{Zaldarriaga:1996xe}
M.~Zaldarriaga and U.~Seljak, \emph{{An all sky analysis of polarization in the
  microwave background}},
  \href{https://doi.org/10.1103/PhysRevD.55.1830}{\emph{Phys. Rev. D}
  {\bfseries 55} (1997) 1830--1840},
  [\href{https://arxiv.org/abs/astro-ph/9609170}{{\ttfamily
  astro-ph/9609170}}].

\bibitem{Kamionkowski:1996ks}
M.~Kamionkowski, A.~Kosowsky and A.~Stebbins, \emph{{Statistics of cosmic
  microwave background polarization}},
  \href{https://doi.org/10.1103/PhysRevD.55.7368}{\emph{Phys. Rev. D}
  {\bfseries 55} (1997) 7368--7388},
  [\href{https://arxiv.org/abs/astro-ph/9611125}{{\ttfamily
  astro-ph/9611125}}].

\bibitem{Lue:1998mq}
A.~Lue, L.-M. Wang and M.~Kamionkowski, \emph{{Cosmological signature of new
  parity violating interactions}},
  \href{https://doi.org/10.1103/PhysRevLett.83.1506}{\emph{Phys. Rev. Lett.}
  {\bfseries 83} (1999) 1506--1509},
  [\href{https://arxiv.org/abs/astro-ph/9812088}{{\ttfamily
  astro-ph/9812088}}].

\bibitem{Feng:2004mq}
B.~Feng, H.~Li, M.-z. Li and X.-m. Zhang, \emph{{Gravitational leptogenesis and
  its signatures in CMB}},
  \href{https://doi.org/10.1016/j.physletb.2005.06.009}{\emph{Phys. Lett. B}
  {\bfseries 620} (2005) 27--32},
  [\href{https://arxiv.org/abs/hep-ph/0406269}{{\ttfamily hep-ph/0406269}}].

\bibitem{McAllister:2008hb}
L.~McAllister, E.~Silverstein and A.~Westphal, \emph{{Gravity Waves and Linear
  Inflation from Axion Monodromy}},
  \href{https://doi.org/10.1103/PhysRevD.82.046003}{\emph{Phys. Rev. D}
  {\bfseries 82} (2010) 046003},
  [\href{https://arxiv.org/abs/0808.0706}{{\ttfamily 0808.0706}}].

\bibitem{Lesgourgues:2011re}
J.~Lesgourgues, \emph{{The Cosmic Linear Anisotropy Solving System (CLASS) I:
  Overview}},  \href{https://arxiv.org/abs/1104.2932}{{\ttfamily 1104.2932}}.

\bibitem{Naokawa:2023upt}
F.~Naokawa and T.~Namikawa, \emph{{Gravitational lensing effect on cosmic
  birefringence}},
  \href{https://doi.org/10.1103/PhysRevD.108.063525}{\emph{Phys. Rev. D}
  {\bfseries 108} (2023) 063525},
  [\href{https://arxiv.org/abs/2305.13976}{{\ttfamily 2305.13976}}].

\bibitem{Sherwin:2021vgb}
B.~D. Sherwin and T.~Namikawa, \emph{{Cosmic birefringence tomography and
  calibration independence with reionization signals in the CMB}},
  \href{https://doi.org/10.1093/mnras/stac3146}{\emph{Mon. Not. Roy. Astron.
  Soc.} {\bfseries 520} (2023) 3298--3304},
  [\href{https://arxiv.org/abs/2108.09287}{{\ttfamily 2108.09287}}].

\bibitem{Galaverni:2023zhv}
M.~Galaverni, F.~Finelli and D.~Paoletti, \emph{{Redshift evolution of cosmic
  birefringence in CMB anisotropies}},
  \href{https://doi.org/10.1103/PhysRevD.107.083529}{\emph{Phys. Rev. D}
  {\bfseries 107} (2023) 083529},
  [\href{https://arxiv.org/abs/2301.07971}{{\ttfamily 2301.07971}}].

\end{thebibliography}\endgroup

\end{document}